\documentclass{article}

\usepackage{arxiv}

\usepackage[utf8]{inputenc} 
\usepackage[T1]{fontenc}    
\usepackage{hyperref}       
\usepackage{url}            
\usepackage{booktabs}       
\usepackage{array}
\usepackage{amsfonts}       
\usepackage{nicefrac}       
\usepackage{graphicx}
\usepackage{subcaption}
\usepackage{adjustbox}
\usepackage[numbers]{natbib}
\usepackage{doi}
\usepackage{amsmath}
\usepackage{amssymb}
\usepackage{algorithm}
\usepackage{algpseudocode}
\usepackage{xcolor}
\usepackage{fancyhdr}
\usepackage{amsthm}
\usepackage{cleveref}
\usepackage{bm}
\usepackage[mathscr]{euscript}
\usepackage{textcomp}
\usepackage{svg}
\usepackage{mathtools}

\DeclareMathAlphabet{\pazocal}{OMS}{zplm}{m}{n}

\usepackage{lscape}		
\usepackage{fancyvrb}

\newtheoremstyle{remarkstyle}  
  {5pt}                        
  {5pt}                        
  {}                           
  {}                           
  {\bfseries}                  
  {.}                          
  { }                          
  {}                           

\theoremstyle{remarkstyle}

\crefname{figure}{Fig.}{Figs.}
\Crefname{figure}{Figure}{Figures}
\crefname{algorithm}{Algorithm}{Algorithms}
\crefname{equation}{Eq.}{Eqs.}

\title{A Multi-Fidelity Bayesian Neural Operator for Mechanics of Spinodal Metamaterial}

\date{} 					

\author{
    \textbf{Pu You},\quad
    \textbf{Hongshun Chen},\quad
    \textbf{Bahador Bahmani}\thanks{Corresponding author. Email: bahador.bahmani@northwestern.edu},\quad
    \textbf{Horacio D. Espinosa}\thanks{Corresponding author. Email: espinosa@northwestern.edu}
    \\[0.5em]
    \textit{\small Department of Mechanical Engineering, Northwestern University, Evanston, IL 60208, USA}
    \\[0.5em]
    \textit{\small Theoretical and Applied Mechanics, Northwestern University, Evanston, IL 60208, USA}
}

\renewcommand{\shorttitle}{A Multi-Fidelity Bayesian Neural Operator for Mechanics of Spinodal Metamaterial}

\hypersetup{
    colorlinks=true,      
    linkcolor=green,      
    urlcolor=blue,        
    citecolor=blue        
}

\begin{document}
\maketitle

\begin{abstract}
    Cellular metamaterials offer a vast design space for tailoring nonlinear mechanical responses, yet exploring this space with conventional modeling approaches is often infeasible or not scalable. To fully exploit their nonlinear behavior for inverse design, it is essential to learn the \textit{full stress–strain response} rather than relying on bulk quantities, motivating the use of neural operators for function-to-function mapping. However, data-driven modeling of nonlinear response for metamaterials is severely constrained by the limited availability of costly experimental data. Here, we propose a Bayesian multi-fidelity deep operator network that aggregates abundant \textit{low-fidelity finite element simulations} with sparse \textit{high-fidelity experimental data} from in-situ nanomechanical experiments on spinodal metamaterials, enabling \textit{heterogeneous information aggregation}. A hybrid Bayesian active learning strategy is introduced to select informative samples by jointly maximizing epistemic uncertainty and geometric diversity of the microstructure, substantially reducing the cost of 3D nonlinear simulations. This approach adaptively trains the low-fidelity operator, which is then augmented by a high-fidelity Bayesian residual learner. We demonstrate that only 22 strategically selected samples from a design pool of 3000 are sufficient to achieve a 84.1\% reduction in MSE compared to the high-fidelity baseline. The framework significantly outperforms single-fidelity baselines, providing superior predictions for full nonlinear stress–strain responses as well as stiffness, strength, and energy absorption. This work provides a robust, data-efficient pathway for the inverse design and constitutive modeling of cellular metamaterials.
\end{abstract}


\section{Introduction}

Advances in additive manufacturing have enabled unprecedented flexibility in the design of materials with tailored mechanical responses that were not achievable using conventional fabrication techniques \cite{jiao2023mechanical}, with applications including—but not limited to energy-absorbing structures \cite{san2020review, kansara2025multi}, lightweight structures \cite{challapalli2021inverse}, and biomedical implants \cite{peng2023machine, vafaeefar2025computational}. This flexibility gives rise to a vast design space, allowing material behavior to be tuned across multiple length scales \cite{zheng2021data, lee2024data, bonfanti2024computational, guo2026functional}. However, increased design freedom is accompanied by heightened complexity in material response \cite{surjadi2025enabling}, often rendering classical constitutive modeling approaches insufficient \cite{fuhg2024review}.

Recent developments in scientific machine learning have opened new opportunities for modeling complex material behavior \cite{zheng2023deep} and for automating the construction of constitutive models directly from data \cite{mozaffar2019deep,flaschel2022discovering, linka2023new, bahmani2024discovering}. While these methods offer remarkable expressive power, their flexibility typically comes at the cost of a large parametric space, which in turn demands orders of magnitude more data than classical hand-crafted models. In the context of material modeling, this data requirement poses a significant challenge, as the preparation of material samples and the execution of mechanical tests are often expensive, time-consuming, and technically demanding \cite{lee2024data}.

Data-driven machine-learning approaches have been widely used to study metamaterials by learning structure–property relationships, often via surrogate models based on multilayer perceptrons, convolutional neural networks, or graph neural networks \cite{kumar2020inverse, yang2020prediction, vlassis2020geometric, gupta2023accelerated, bastek2023inverse, thakolkaran2025experiment}. Most existing methods, however, are formulated as regression models rather than learning the underlying partial differential equation (PDE) operator governing the microstructural response, limiting their generalization across loading conditions, boundary conditions, and discretizations. Recent operator learning frameworks address this limitation by directly approximating solution operators, allowing more flexible surrogate models \cite{lu2019deeponet,li2020fourier,bhattacharya2021model, he2023novel, jin2025characterization, bahmani2025resolution, liu2025towards}. A further challenge is data scarcity, especially for spinodal metamaterials and other materials with complex microstructures, where high-fidelity data generation is costly. Multi-fidelity frameworks \cite{peherstorfer2018survey,meng2021multi} offer a natural remedy by integrating heterogeneous data sources, such as low- and high-fidelity simulations \cite{cheung2024multi, zhu2025frequency, lindsey2025chimes} or combinations of computational and experimental data \cite{fare2022multi, noh2026performance}.

A representative class of materials where these modeling challenges are particularly acute is spinodal metamaterial. Conventional unit cells that constitute metamaterial, e.g., truss and shell-based periodic lattices, are usually sensitive to imperfection or mechanically inefficient in utilizing materials due to the single damage band \cite{gunther2023experimental}. In contrast, spinodal architectures are porous structures with non-periodic and intertwined topologies and are known for their reduced sensitivity to fabrication-induced imperfections \cite{hsieh2019mechanical, zhang2021mechanical, surjadi2025enabling}. Remarkably, even with a small number of design parameters—such as three cone-angle parameters with one relative density—spinodal architectures can generate an effectively infinite family of smooth and distinct topologies \cite{kumar2020inverse}. Although significant progress has been made in the generation and geometric control of spinodal topologies \cite{ montes2021accelerating, oommen2022learning, hu2022accelerating,wang2023machine,  oommen2024rethinking}, a comprehensive understanding of their nonlinear mechanical behavior remains an active area of research, particularly with respect to their fracture properties \cite{Chen2026beyond}, damage resistance \cite{hsieh2019mechanical, zhang2021mechanical}, recoverability \cite{portela2020extreme, anandan2025functionally}, and anisotropic behavior \cite{senhora2022optimally, viet2023directional, dhulipala2025curvature}. Its extremely complex 3D microstructure also hinders the generation of large amounts of high-fidelity nonlinear response data.

To mitigate the challenges of data scarcity and geometric complexity in spinodal metamaterials, we proposed a multi-fidelity (MF) modeling framework in which data from low-fidelity but computationally inexpensive simulations are leveraged to constrain and inform high-fidelity, data-driven surrogate models. By systematically integrating experimental observations with auxiliary simulation data, the proposed approach seeks to reduce data requirements while retaining predictive accuracy and physical relevance. We treat the experimental data as ground-truth, high-fidelity (HF) information, while acknowledging the presence of measurement noise and fabrication-induced imperfections. The low-fidelity (LF) data consist of finite element simulations of the spinodal microstructure. These simulations are further considered low-fidelity due to the use of simplified constitutive models for the local material behavior, whose parameters are calibrated using experimental data from standard tests.

\begin{figure*}[!htbp]
    \centering
    \includegraphics[width=1\linewidth]{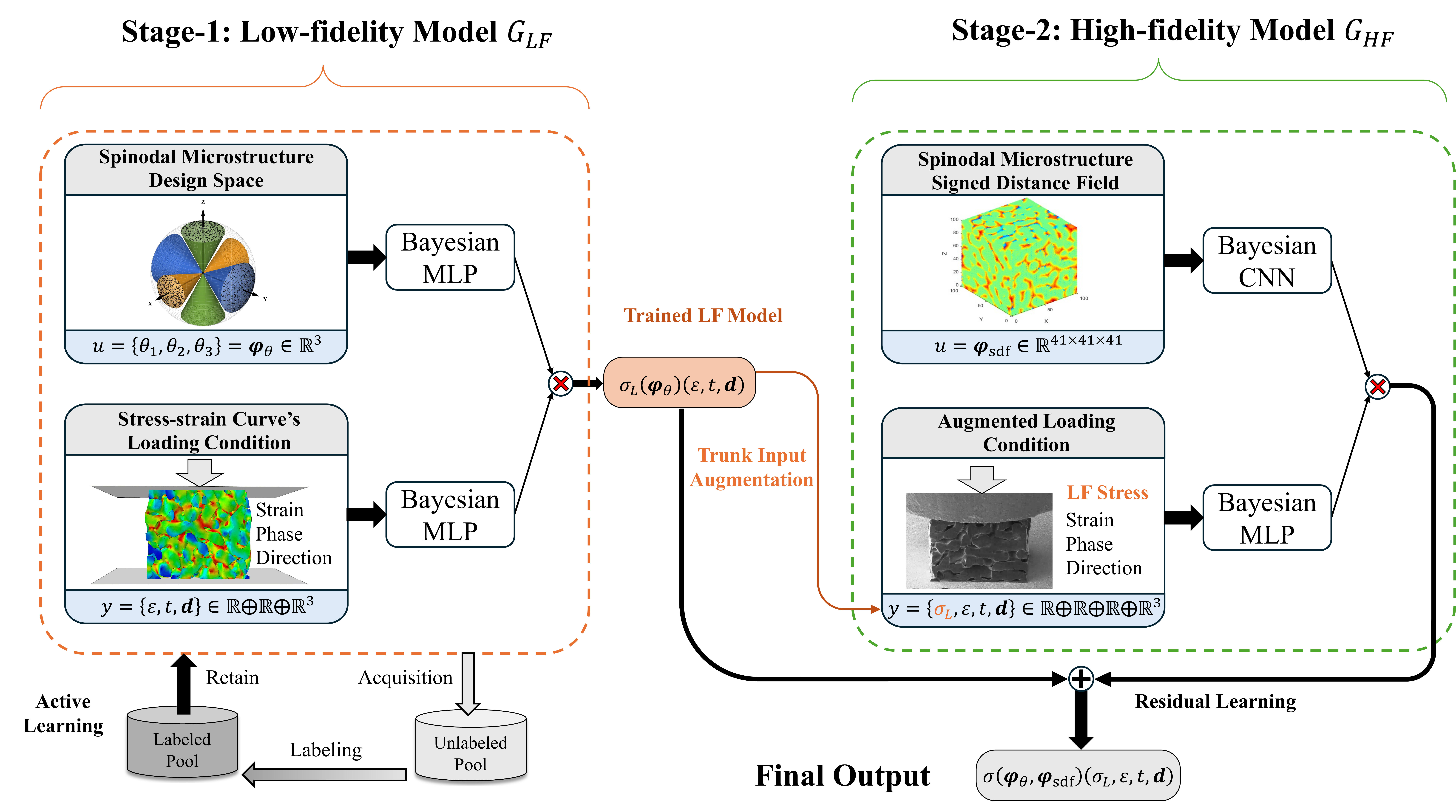}
    \caption{The MF operator learning framework. The LF model acts as a frozen baseline, while the HF residual model utilizes a 3D CNN to map the Signed Distance Field to the simulation-experiment discrepancy.}
    \label{Whole_architecture}
\end{figure*}

 We leverage a classical MF surrogate modeling strategy in which predictions from a LF model are used as augmented inputs to inform a HF model \cite{kennedy2000predicting, perdikaris2017nonlinear, lu2022multifidelity}. We construct both the LF and HF surrogate models using the DeepONet framework \cite{blundell2015weight, lu2019deeponet} to enable fast and differentiable predictions. Uniquely, the HF branch incorporates a 3D Convolutional Neural Network (CNN) to directly process the full 3D Signed Distance Field (SDF) of the microstructure. Furthermore, the operator is extended to a Bayesian setting to estimate uncertainty \cite{blundell2015weight}. To maximize data efficiency, we introduce a hybrid active learning strategy that iteratively selects LF simulation samples by maximizing a composite score of epistemic uncertainty (exploitation) and geometric diversity (exploration).  Finally, our framework builds a MF Bayesian DeepONet that directly predicts stress–strain curves under uniaxial loading in three orthogonal directions, as shown in Figure \ref{Whole_architecture}. This framework allows for the development of a highly accurate, probabilistic surrogate model even when HF experimental data is exceptionally scarce.

To the best of our knowledge, this is the first MF framework that aggregates LF simulations with HF experimental data to model full three-dimensional, highly nonlinear stress responses arising from plasticity and contact in spinodal metamaterials. With only 20 HF samples and 22 strategically selected LF samples, the proposed approach consistently outperforms single-fidelity models, reducing the MSE by 86.0\% and 84.1\% compared to the LF and HF models, respectively. This work may motivate further research to leverage the proposed framework for improved understanding of the complex mechanics of such metamaterials and for goal-oriented design.

The outline of this paper is as follows: Section \ref{section_results} first introduces the design space of spinodal microstructure, and then presented the generation of the MF database (comprising finite element simulations and in-situ nanomechanical experiments). Based on the propped model, we discussed the results by demonstrating the efficiency of the active operator learning strategy and the superior predictive accuracy of the MF model compared to single-fidelity approaches. After that, we provide a summary of this work and our discussion in future work. Section \ref{section_methods} presents the fundamental mathematical formulation of MF and BNN operator learning . Followed by that is the implementation details of hybrid active learning strategy in LF model training and SDF-based residual learning in HF model.

\section{Results and Discussion}
\label{section_results}
\subsection{Spinodal Microstructure with Tunable Topology}
The problem at hand is to collaborate simulations with in-situ experiments to construct an efficient surrogate model for predicting the nonlinear mechanical properties of spinodal metamaterials. Spinodal architectures mimic the stochastic bicontinuous topologies formed when a homogeneous solution undergoes phase separation into two distinct phases, and this process is governed by the Cahn-Hilliard equation \cite{grant1993spinodal}. Solving the Cahn-Hilliard partial differential equations (PDEs) directly for large volumes is computationally expensive. Cahn et al. demonstrated that the resulting phase field can be effectively approximated by the superposition of a large number of standing waves, known as Gaussian Random Fields (GRF) \cite{kumar2020inverse}. For an isotropic structure, the mathematical formulation of the field function $\varphi(\mathbf{x})$ is given by:

\begin{equation}
    \varphi(\mathbf{x}) = \sqrt{\frac{2}{N}} \sum_{i=1}^{N} \cos(\beta \mathbf{n}_{i} \cdot \mathbf{x} + \gamma_{i}), \quad \text{with } \mathbf{n}_i \sim \mathcal{U}(\mathbb{S}^2), \, \gamma_i \sim \mathcal{U}([0, 2\pi))
\end{equation}

where $\mathbf{x}$ is the 3D spatial coordinate, $N$ is the number of superimposed waves, and $\beta$ is the wavenumber controlling the characteristic length scale. The direction wavevectors $\mathbf{n}_i$ are uniformly distributed on the unit sphere surface $\mathbb{S}^2$, and the phase shifts $\gamma_i$ are uniformly distributed between $[0, 2\pi)$. In this formulation, $\beta$ is a fixed constant ensuring structural periodicity, while $\mathbf{n}_i$ and $\gamma_i$ act as stochastic variables determining the specific morphology.

To separate the material phases, a level-set function $\chi(\mathbf{x})$ is applied to the continuous field $\varphi(\mathbf{x})$. For solid-based spinodal structures, the domain is defined as:

\begin{equation}
    \chi(\mathbf{x}) = 
    \begin{cases} 
    1, & \text{if } \varphi(\mathbf{x}) \leq \varphi_{0} \\
    0, & \text{if } \varphi(\mathbf{x}) > \varphi_{0} 
    \end{cases}
\end{equation}

\noindent where $\varphi_0$ is a threshold parameter derived from the desired relative density (porosity) $\bar{\rho}$. The relationship is given by $\varphi_{0} = \sqrt{2} \text{erf}^{-1}(2\bar{\rho}-1)$. Previous studies indicate that maintaining bi-continuity typically requires $\bar{\rho} > 0.159$ \cite{soyarslan20183d}. Within the same level-set,  extracting the iso-surface at $\varphi(\mathbf{x}) = \varphi_0$ we can also generate shell-based structures.

The stochastic nature of the spinodal microstructure is governed by the distribution of wave vectors $\mathbf{n}_i$. By constraining the sampling domain of $\mathbf{n}_i$, diverse anisotropic topologies can be generated. Instead of sampling uniformly from the entire unit sphere, we can restrict $\mathbf{n}_i$ using three defining cone angles $\{\theta_1, \theta_2, \theta_3\}$ relative to the Cartesian basis vectors $\{\hat{\mathbf{e}}_1, \hat{\mathbf{e}}_2, \hat{\mathbf{e}}_3\}$:

\begin{equation}
    \mathbf{n}_{i} \sim \mathcal{U}\left( \left\{ \mathbf{k} \in \mathbb{S}^{2} : (|\mathbf{k} \cdot \hat{\mathbf{e}}_{1}| > \cos\theta_{1}) \oplus (|\mathbf{k} \cdot \hat{\mathbf{e}}_{2}| > \cos\theta_{2}) \oplus (|\mathbf{k} \cdot \hat{\mathbf{e}}_{3}| > \cos\theta_{3}) \right\} \right)
\end{equation}
where $\mathbf{k}$ is the confined direction wavevector, and $\oplus$ is the operator symbol for combining three spaces.

\begin{figure*}[!htbp]
    \centering
    \includegraphics[width=1\linewidth]{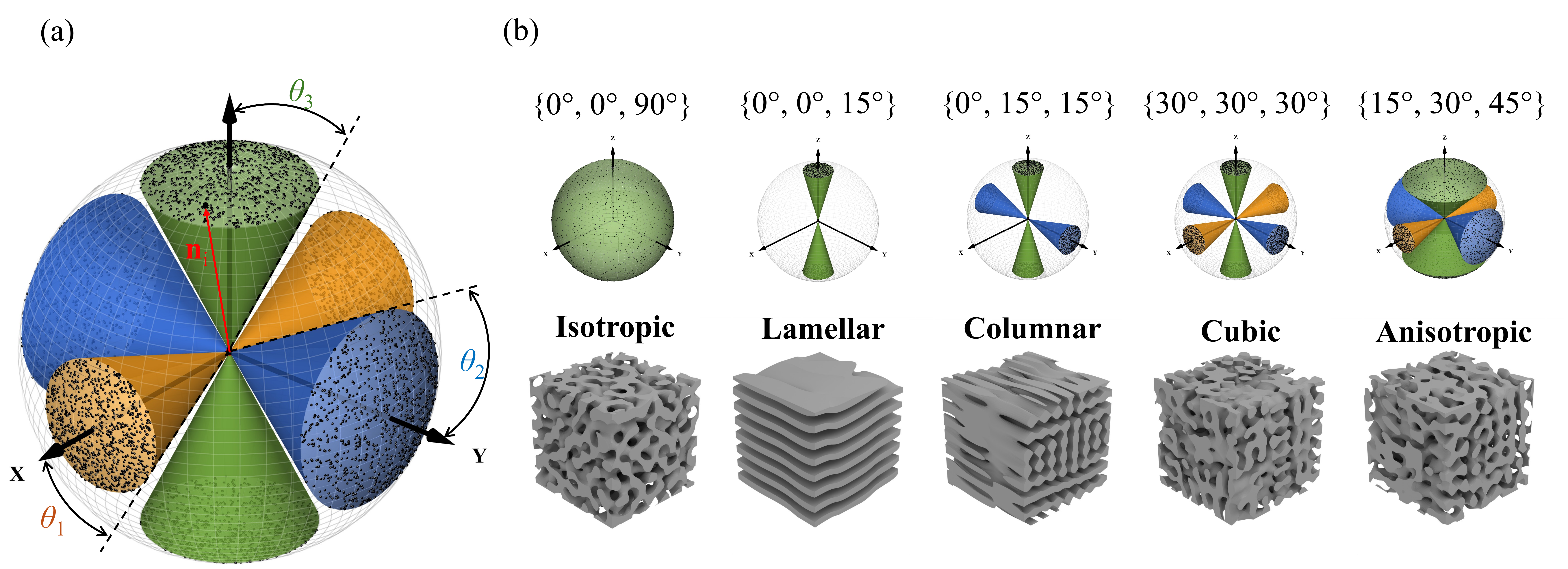}
    \caption{Schematic diagrams of design angles and different microstructure. (a) Illustration of cone angle constraint for random vectors. (b)Five typical microstructures generated from different design angles}
    \label{Design_space}
\end{figure*}

As shown in Figure \ref{Design_space}, these angles define six spherical caps within six cones, and the random vectors should be confined within these regions. To ensure structural continuity using this definition, the angles are restricted to $\theta_i \in \{0^\circ\} \cup [15^\circ, 90^\circ]$. By varying these angles, distinct microstructural classes can be obtained: $\{0, 0, \theta_1\}$ can generate lamellar structures; $\{0, \theta_2, \theta_1\}$ yields columnar structures; and $\{\theta_3, \theta_2, \theta_1\}$ produces general anisotropic cubic structures. Five typical microstructures are shown in Figure \ref{Design_space} (b). In this work, we fix the relative density at $\bar{\rho}=0.45$ and investigate the mechanical response of solid-based spinodal microstructures driven by variations in these three design angles. All structures were generated using the open-source GibbonCode MATLAB package \cite{GibbonCode}.

\subsection{Low-Fidelity Data: Finite Elements Simulation}
\label{sec:lofi_generation}
To construct the LF database, we first partitioned the design space into three distinct topological groups based on the cone angle parameters: lamellar $(0,0,\theta_1)$, columnar $(0,\theta_2,\theta_1)$, and general anisotropic $(\theta_3,\theta_2,\theta_1)$. For each group, we employed Latin Hypercube Sampling to select 1000 unique design parameter sets. Thus, there are 3000 sets of angles in the pool, and the angles distribution is shown in Supplementary Figure SI.1. For specific angles, the geometric models were generated and saved as STL files using MATLAB R2024a with a fixed random seed (seed=1) to ensure reproducibility. The cube has 7.5 unit cells (or 15$\pi$) in each direction, and 1000 waves were superimposed to produce GRF. Subsequently, the open-source library \texttt{fTetWild} \cite{hu2020fast} was utilized to perform robust tetrahedral mesh generation using default parameters. An automated Python pipeline was developed to generate the simulation input files for ABAQUS 2024. During the subsequent active operator learning stage, batches of angle sets were selected for labeling via FEM simulations using the ABAQUS/Explicit solver. Detailed FE setups are shown in the Supplementary Note S1. All computational tasks were performed on the Quest supercomputing cluster at Northwestern University.

\subsection{High-Fidelity Data: In-situ Nanomechanical Experiment}
\label{sec:hifi_experiments}

High-fidelity data were obtained through in-situ nanomechanical testing following established protocols \cite{jin2025characterization}. Stochastic spinodal microstructures were fabricated using high-resolution Two-Photon Lithography (Nanoscribe GmbH) and prepared via critical point drying to preserve morphological integrity. Uniaxial compression tests were subsequently conducted along three principal directions inside a Scanning Electron Microscope (SEM) using a displacement-controlled nanoindentation platform (Alemnis AG). The acquired load-displacement data were corrected for system compliance and converted into engineering stress–strain curves. The final HF dataset consists of 20 microstructures (60 curves) for model training, while 5 microstructures (15 curves) were reserved for testing. Comprehensive details regarding sample fabrication, testing parameters, and data processing are provided in Supplementary Note S2.

\subsection{Active Operator Learning Using a Hybrid Strategy}
The training of the MF model consists of two steps: first, training the LF model using active operator learning (AOL) with a hybrid strategy, and then performing residual learning on the HF data based on the learned LF model. The first training stage is based on LF data. The goal of this stage is to explore a sufficiently large design space so that the LF model can learn mechanical response characteristics from various microstructures patterns, such as hardening and softening. Using the AOL strategy described in the algorithm \ref{alg:hybrid_al}, model starts learning from two microstructures and then selects new samples for labeling based on the hybrid AOL strategy. During the batch labeling process, ten new samples are added from the sample pool at each iteration. Each training was repeated five times using 5 different random seeds, and the model with the smallest loss was used. The details of training and loss history are provided in the Supplementary Figure SI.3.

\begin{figure*}[!htbp]
    \centering
    \includegraphics[width=1\linewidth]{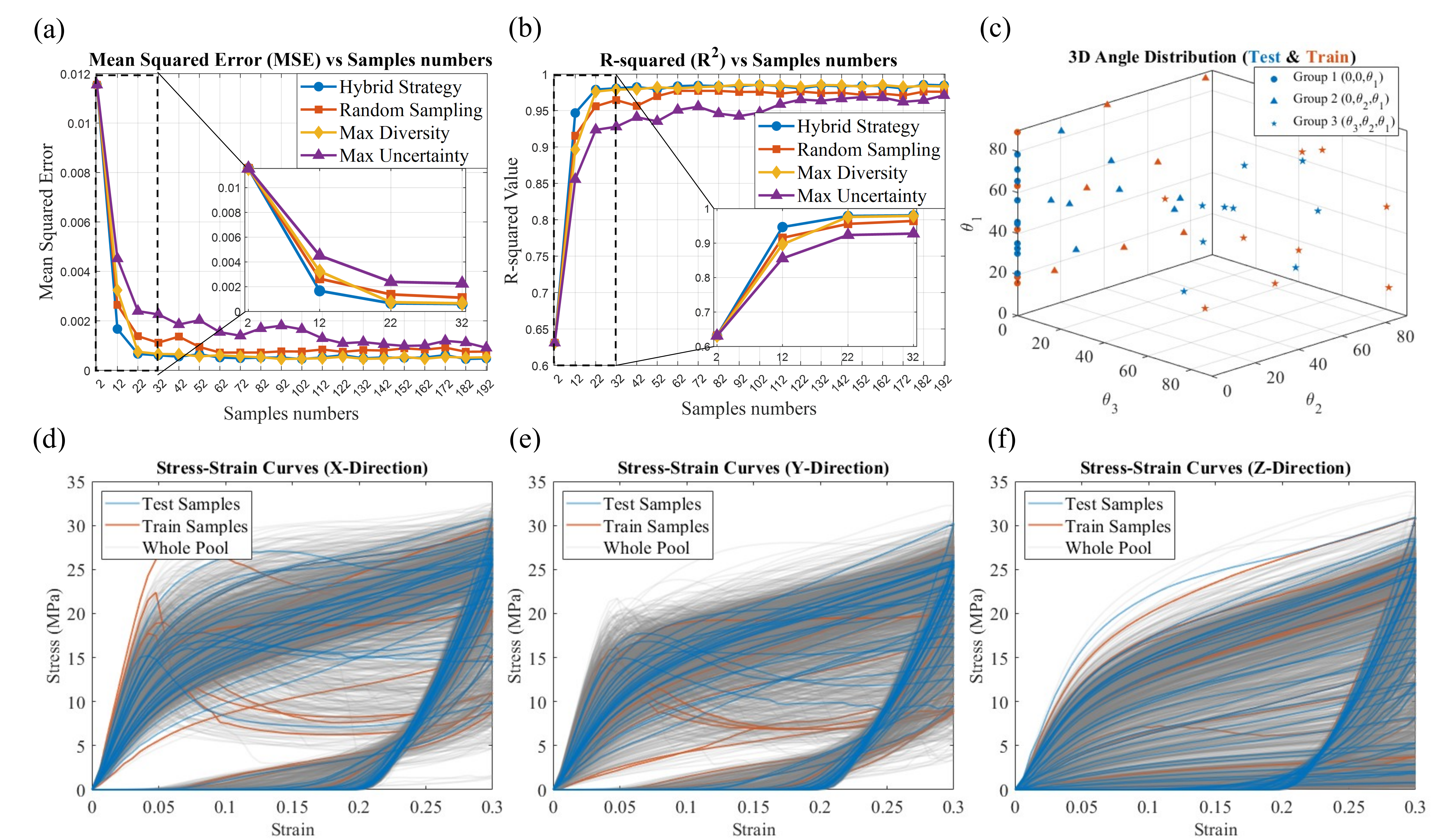}
    \caption{\textbf{(a-b)} MSE and R-squared vs samples number in four different schemes during active learning. The hybrid strategy is the proposed scheme, combining two terms; random sampling means adding samples without any bias; while the other two strategies correspond to cases where the hybrid strategy degenerates into only including one term. \textbf{(c-f)} Angles and curves distribution when there are only 22 training samples. The orange lines and dots represent 22 training samples, the blue represent 30 test samples, and the gray lines represent stress–strain curves obtained from all 3000 samples in our pools.}
    \label{Result_LF}
\end{figure*}

Next, the LF model was tested on 30 unseen microstructures that were randomly selected from the initial pool of 3000 samples(each of the 3 categories contains 10 samples). Predictions were made for all 30 test samples. Each sample made 50 separate forward predictions to obtain the mean and variance of stress–strain curves. 
We implemented three other strategies for new sample acquisition based on: 1) random sampling; 2) only uncertainty metric; and 3) diversity metric. 

Figure \ref{Result_LF} (a-b) shows the MSE and R-squared values of the trained model's prediction on the test samples when using different number of training samples. It demonstrates that the hybrid strategy provides the fastest convergence with smallest number of samples among all strategies. In particular, the maximum uncertainty (purple line) leads to the worst result, even worse than random sampling (red line). Under maximum uncertainty strategy, the model always tends to select new samples with angles $(0, 0, \theta_1)$. This occurs because lamellar microstructures exhibit more complex stress responses, leading to persistently higher predictive uncertainty. Thus focusing only on the uncertainty may lead to over-concentration on narrow design angles region, thus resulting in a lack of global exploration capability for different angles patterns. The second worst strategy is random sampling.
Following that, the maximum diversity (yellow line) is the second best, and is significantly better than random sampling. The hybrid strategy, combining both diversity and uncertainty metrics, (blue line) stands out, since it further improves learning efficiency via efficiently balancing global exploration and exploitation to identify samples with high uncertainty.

Another noteworthy point is that the hybrid strategy achieved nearly saturated MSE and R-squared using only 22 samples. To reveal the underlying mechanism, we plot the curve distributions of the training sample, test sample, and whole pool sample, respectively, as well as the angle distributions of them, as shown in Figure \ref{Result_LF} (c-f). The orange lines have diverse patterns and cover most of the area, encompassing various types of stress–strain behaviors such as hardening and softening. Furthermore, the angle distribution is sufficiently uniform, with a reasonable distribution and division across groups 1, 2 and 3 (correspond to lamellar, columnar and general anisotropic microstructures). This indicates that through active operator learning with  hybrid strategy, model can represent most of response patterns with only 22 samples to achieve sufficiently good training performance.

\begin{figure*}[!htbp]
    \centering
    \includegraphics[width=1\linewidth]{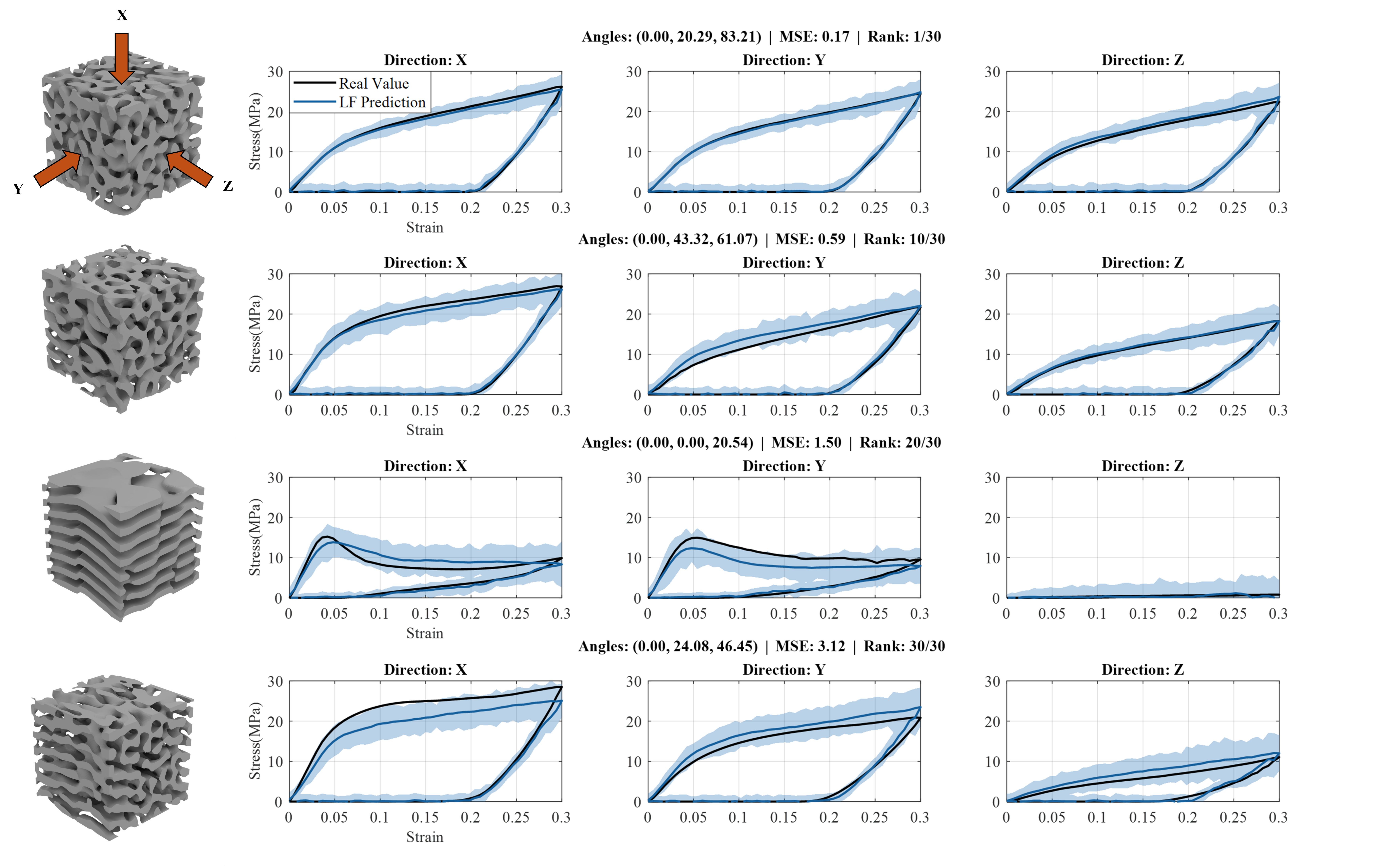}
    \caption{The LF surrogate model's predicted stress–strain curves for 4 samples. From top to bottom, they are: nearly isotropic, general anisotropic, lamellar, and columnar microstructure.}
    \label{Section4-2fig4}
\end{figure*}

Based on the trained LF model, predictions were made for all 30 test samples. We then ranked the MSE of the 30 samples from lowest to highest (meaning from best fit to worst fit) and selected 4 samples for display, as shown in Figure \ref{Section4-2fig4}. Specifically, the first sample exhibits the closest characteristics to isotropic behavior, and the curves in the three directions are the most consistent. The second sample shows a general anisotropic microstructure. The third sample is a typical lamellar microstructure, with the curves in the X and Y directions exhibiting significant softening behavior due to the buckling of shell-like microstructure, while the compression curve in the Z direction has low stress level attributed to the lack of structural connection and supports. The fourth sample is a columnar microstructure, with the highest stress direction in the X direction.

In addition, PICP (formula in Equation \ref{PICP}) was calculated for the curves in three directions for each of the four samples, as shown in Table \ref{PICP_LF}. In the best-fitting case, all stress points fall within the 95\%-confidence interval. For samples ranked 20/30, 100\% of the stress points in two directions fall within this interval, while over 95\% fall within the interval in the remaining direction. In the worst-case scenario, 100\% of the stress points fall within the interval in two directions, and 84\% in the remaining direction. Here, the PICP metrics also provide the same conclusion as mentioned above: the best prediction performances are for the nearly isotropic microstructure, as its features are the simplest and easiest to learn. The worst prediction performances are for the columnar and lamellar structures, due to their potentially greater uncertainty (arising from complex physical behavior such as buckling and self-contact in beam and shell-like microstructure).

\begin{table}[ht]
    \centering
    \caption{Prediction Interval Coverage Probability (PICP) for 4 Selected Samples}
    \label{tab:picp_results}
    \begin{tabular}{lcccc}
        \toprule
        & \textbf{Rank 1} & \textbf{Rank 10} & \textbf{Rank 20} & \textbf{Rank 30} \\
        \textbf{Direction} & $(0^\circ, 20.29^\circ, 83.21^\circ)$ & $(0^\circ, 43.32^\circ, 61.07^\circ)$ & $(0^\circ, 0^\circ, 20.54^\circ)$ & $(0^\circ, 24.08^\circ, 46.45^\circ)$ \\
        \midrule
        X-Direction & 100.00\% & 100.00\% & 100.00\% & 84.16\% \\
        Y-Direction & 100.00\% & 98.02\%  & 95.05\%  & 100.00\% \\
        Z-Direction & 100.00\% & 100.00\% & 100.00\% & 100.00\% \\
        \bottomrule
    \end{tabular}
    \label{PICP_LF}
\end{table}

\subsection{Multi-fidelity Operator Learning and Model Comparison}

The second training step is for HF model (the green box in the Figure \ref{Whole_architecture}) based on the pre-trained LF model and experimental data. This step utilizes the previously trained LF model to augment the trunk network input, while simultaneously learning the residual between LF and HF model. In this training, 20 samples' HF data were used for training, and the remaining 5 samples were used for testing. We show 4 metrics for the training process, including learning rate, KL divergence, loss, and MSE, as detailed in  Supplementary Figure SI.4. 

\begin{figure*}[!htbp]
    \centering
    \includegraphics[width=1\linewidth]{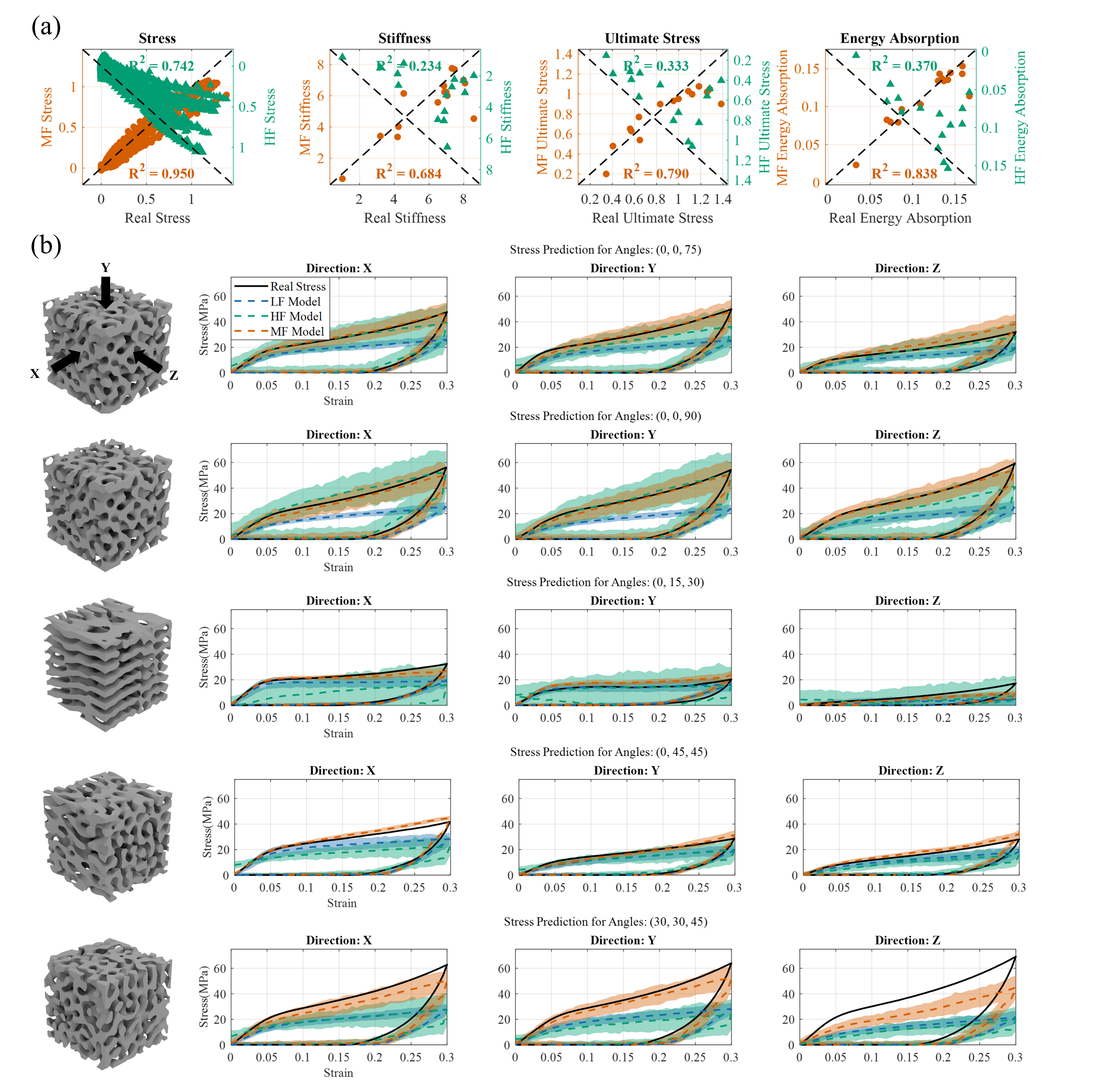}
    \caption{\textbf{(a)} Mechanics properties prediction using MF surrogate model and HF surrogate model. Orange dots represent the MF prediction, and green dots represent the HF prediction. \textbf{(b)} Stress prediction of MF, LF and HF surrogate model on 5 unseen samples. The LF model represents using only FEM simulation data to predict the stress–strain responses of unseen microstructure; the HF model represents using only experimental data to do prediction; and the MF model is the comprehensive method of the framework proposed in this work, which combines both LF and HF models}
    \label{Result_MF}
\end{figure*}

For this training, we expect the MF model to learn real mechanical characteristics from various microstructure patterns with the help of LF model. Here, we compared the prediction results (i.e., stress, stiffness, ultimate stress, and energy absorption) of the two models in Figure \ref{Result_MF} (a): the HF model uses only HF data, while the MF model uses the full proposed framework. For the three properties derived from stress (stiffness, ultimate stress, and energy absorption), the MF model significantly outperforms the standalone HF model. Most of the MF predictions are near the axis of ground-truth, while the HF predictions are highly dispersed, as shown in Figure \ref{Result_MF} (a). This means that the MF model far surpasses the standalone HF model due to the sparse HF training data set of 20 in deeper curve patterns learning.

\begin{figure*}[!htbp]
    \centering
    \includegraphics[width=1\linewidth]{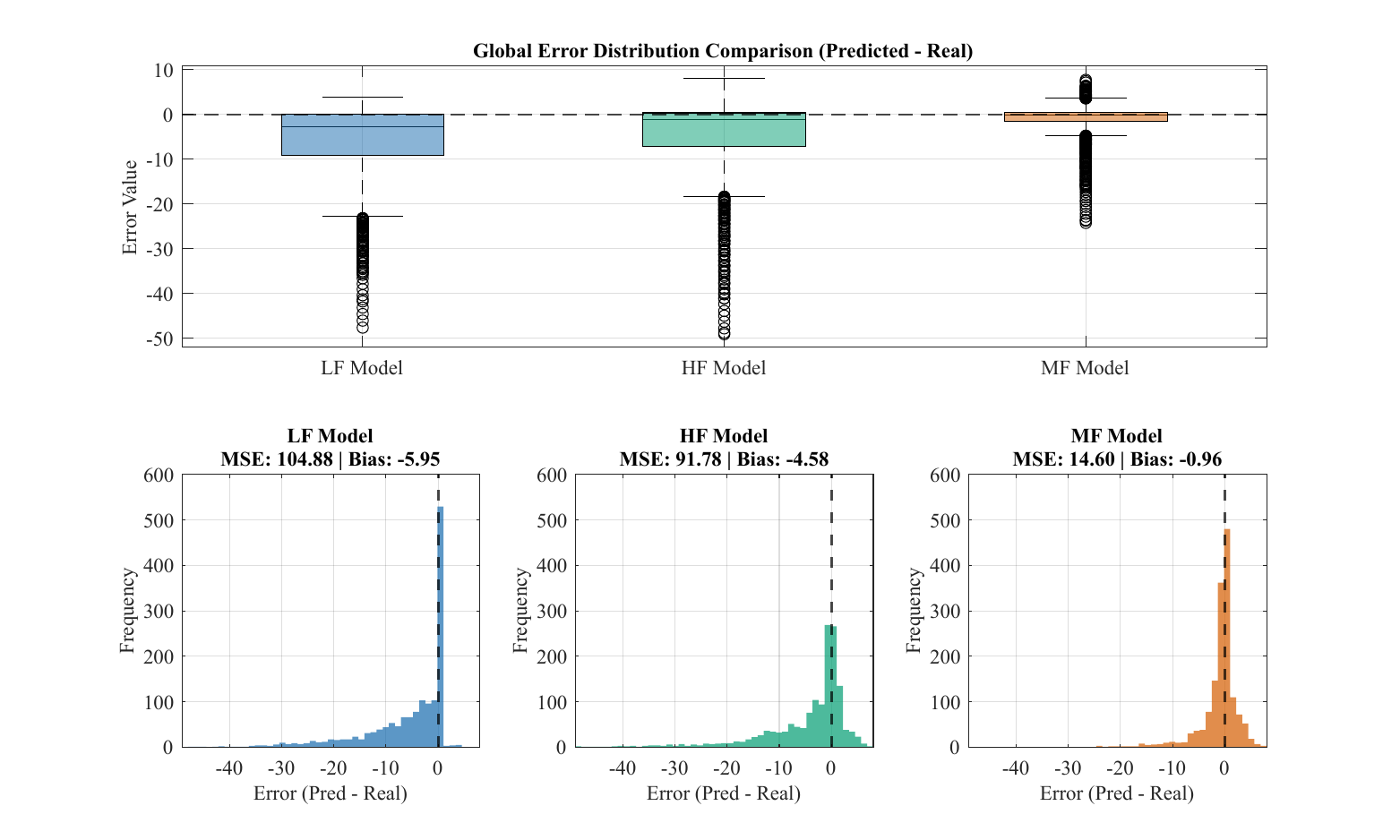}
    \caption{Stress prediction error of MF, LF and HF surrogate model on 5 unseen samples. The upper is a box plot showing the error spread, and the lower is the histogram showing the error distribution.}
    \label{Section4-3fig4}
\end{figure*}

To further examine the performance of the surrogate models from different data sources, we used LF, HF, and MF models, respectively, to predict the stress–strain responses based on 5 unseen microstructures as shown in Figure \ref{Result_MF} (b). The LF model's prediction tend to have a very narrow uncertainty band (blue), because most of factors in the simulation are deterministic, and the main uncertainty arises from the stochastic nature of the spinodal microstructure. The HF model's results have a very wide uncertainty band (green) due to the insufficient training data. In comparison, the MF model exhibits higher accuracy and a smaller band. We also show the error distribution plot in Figure \ref{Section4-3fig4}, where the MF model's errors are much smaller and more concentrated than the standalone LF and HF models. In comparison quantitatively, the MF modeling outperforms the other two, with the MSE reduced by 86.0\% and 84.1\% compared to the LF and HF models, respectively.

With only 20 HF samples' experimental data, HF model training is unreliable under such sparse data conditions. The explanation for this phenomenon is: in such sparse HF data, the HF model training without the help of LF data can only focus on the superficial features of the stress curve, failing to uncover deeper features, which are often masked by experimental uncertainties (e.g., microstructure manufacturing, experimental setup, material aging characteristics, etc.). In the generation of LF FEM data, these settings are often deterministic, thus allowing for the training of a more reliable LF model. However, the LF model cannot represent the actual physical situation, so it also has significant gap between LF and HF data. Furthermore, the MF model with LF model augmentation only needs to learn the residual between the LF and HF data, rather than the complete curve. Therefore, this learning characteristic makes MF more stable and allows it to utilize LF data more efficiently to compensate for the high uncertainty of the HF data.

\subsection{Conclusion}
\label{section_result}

In this work, we proposed a Multi-Fidelity Bayesian Operator Learning framework designed to predict the full-field homogenized non-linear mechanical response of spinodal architected materials. By integrating abundant, low-cost Finite Element simulations with sparse, high-cost in-situ experimental data, we addressed the challenges of experimental data scarcity and idealized low-fidelity simulation data incapable of accounting for all contributing factors in data-driven mechanics.

Through a two-stage training process, the main points of the proposed framework includes:  

\begin{enumerate}
    \item A Hybrid Active Operator Learning strategy for the low-fidelity surrogate model was proposed. By balancing uncertainty-based exploitation with diversity-based exploration, the active learning algorithm achieved a matured prediction capability using \textbf{only 22 samples} selected from a pool of 3000, demonstrating exceptional data efficiency compared to random sampling or single-metric strategies.
    \item The MF model learned the discrepancy between the idealized simulations and the physical experiments. The LF model plays the role in the trunk net input augmentation and residual learning. By incorporating the full 3D geometry via SDF into a CNN-based branch net, the MF model successfully captured manufacturing-induced geometric uncertainties that simple design parameters could not represent.
    \item The results indicate that the MF model significantly outperforms single-fidelity approaches. Specifically, the MF model provided superior predictions for stress-derived mechanical properties, i.e., stiffness, ultimate stress, and energy absorption. The model trained solely on sparse high-fidelity data exhibited large uncertainty bands and poor generalization due to limited data size, while the MF model leveraged the global trends learned by the LF baseline to produce tighter confidence intervals and higher accuracy. 
\end{enumerate}

This framework establishes a robust pathway for leveraging MF-BNN-DeepONet in the characterization of complex architected materials, offering a scalable solution for inverse design and constitutive modeling where experimental resources are limited. In this context, it represents an alternative to our prior neural operator work \cite{jin2025characterization}. Our future work will build on this MF framework for inverse design of spinodal microstructures, with the expectation of developing accurate and data-efficient inverse models.

\section{Methods}
\label{section_methods}

\subsection{Multi-Fidelity Bayesian Operator Framework}
To predict the stress–strain response of spinodal microstructures with quantified uncertainty and active learning, we propose a Multi-Fidelity Bayesian Operator Learning framework (MF-BNN-DeepONet). This approach maps the input function space  $\mathcal{U}$  (design parameters and geometry) and the coordinate space  $\mathcal{Y}$  (loading conditions) to the output stress via operator $G: \mathcal{U} \times \mathcal{Y} \to \mathbb{R}$. The framework combines LF simulation data with HF experimental data through a two-step training process. The following subsections will introduce the detailed framework of MF-BNN-DeepONet.

\subsection{Multi-Fidelity Operator Learning}
\label{sec:MFO}
This subsection introduces MF modeling techniques to leverage multiple data types. Herein, for spinodal microstructure, we use simulation results to represent LF data and in-situ experimental results to represent HF data. LF data are inexpensive and have fewer uncertainties, but they also deviate from the real physical world (e.g., fabrication defects, mesh artifacts, insufficient or inaccurate representation of physical behaviors such as micro-cracking, contact, buckling, plasticity, etc.). HF experimental data is expensive to acquire and time-consuming. In addition, the sample manufacturing and experimental processes involve unpredictable and undiscovered factors, yet it accurately accounts for the real physical behaviors. In this context, we use an MF operator learning framework to leverage the advantages of MF source data. 

The Vanilla DeepONet with branch and trunk networks, respectively, is the core architecture in this work, as shown in Figure \ref{Result_LF}. Its purpose is to map the structural information to full stress–strain curves. The DeepONet \cite{lu2019deeponet} is a typical scientific machine learning architecture designed to learn operators and mappings between infinite-dimensional function spaces \cite{lu2021learning}. The DeepONet architecture accomplishes this via two sub-networks: a "branch" net that encodes the input parameters and a "trunk" net that encodes the output function's coordinates, with their latent vectors combined to approximate the operator \cite{lu2019deeponet, lu2021learning, lu2022comprehensive}. This is ideally suited for the data-driven modeling of spinodal mechanical properties, which involves learning the mapping from a set of design parameters or geometry information (input function) to the full stress–strain response (output function). Due to its powerful ability, it has wide potential in the training of PDE problems \cite{lu2021learning, wang2021learning, howard2023multifidelity, franco2023mesh, he2024geom} and engineering applications \cite{RN16, RN15, RN18}. Furthermore, in the present work, we also introduced a Bayesian neural network  (as the weights distribution in Figure \ref{Result_LF}) in the MF model to quantify the uncertainty, the formulation of which will be discussed in the Section \ref{sec:BNN}.

\begin{figure*}[!htbp]
    \centering
    \includegraphics[width=0.8\linewidth]{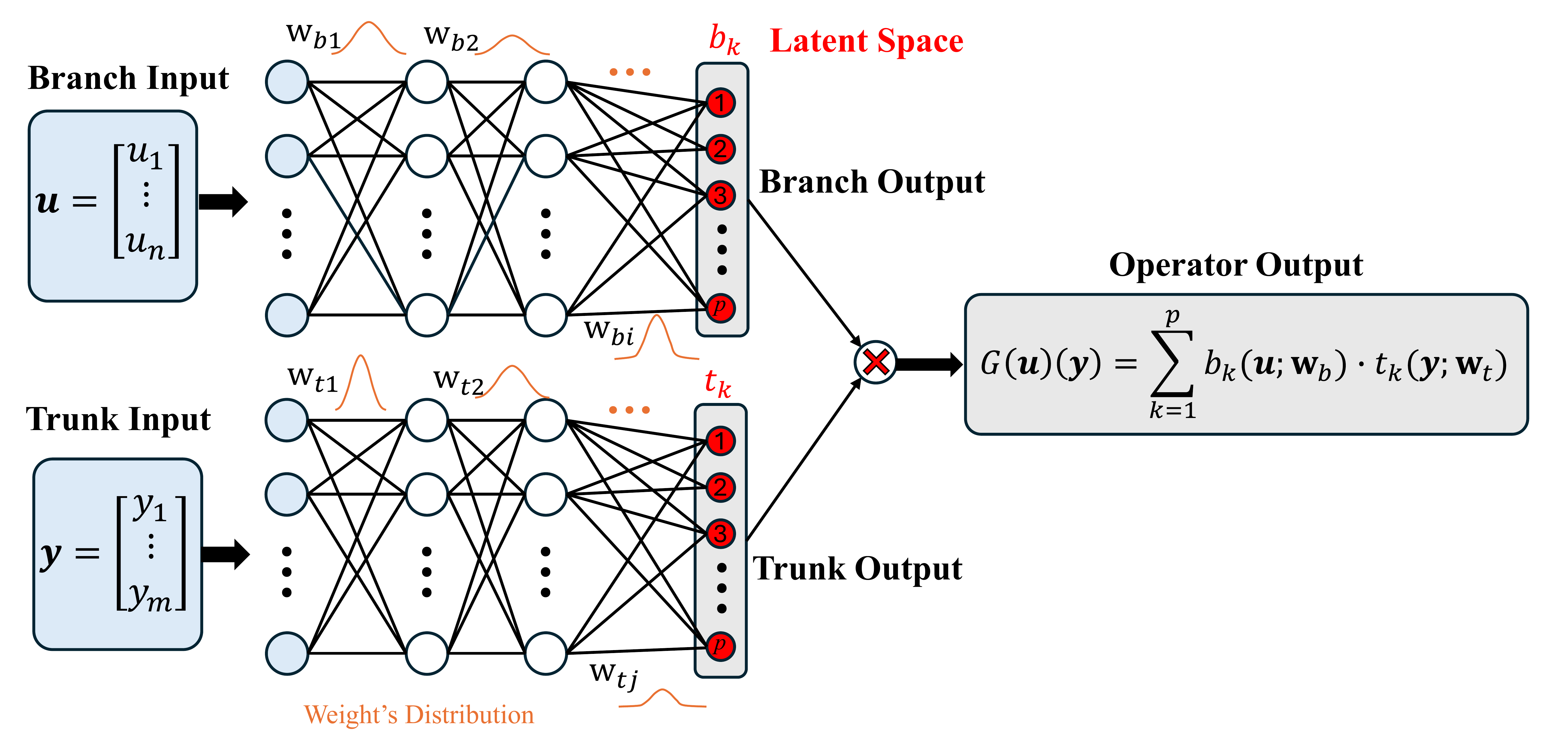}
    \caption{Architecture of the Bayesian DeepONet. The Branch network encodes the input geometry information, while the Trunk network encodes the query coordinates. BNN assigned each weight a distribution to enable uncertainty quantification. }
    \label{BNN_DeepONet}
\end{figure*}
Figure \ref{Whole_architecture}  shows the MF framework used in this work which consist of a two-step training.  The first step is training the LF model (orange box of Figure \ref{Whole_architecture}), while the second step exploits the pretrained LF model to augment the trunk input and to modify the training output as residual in HF model (green box of Figure \ref{Whole_architecture}). Specifically, we pass the HF data's input through the LF model to obtain a LF prediction, which was then incorporated as augmented input to the HF trunk network. The learning target is no longer the original output label, but rather the residual between the LF model output and the HF model label. Accordingly, we represent the true MF response \(G_{MF}\) as the superposition of the LF prediction \(G_{LF}\) and a residual term $\delta_{HF}$ learned from HF model:
\begin{equation}
    G_{MF} = \underbrace{G_{LF}(u)(y)}_{\text{Frozen Prediction}} + 
    \underbrace{\delta_{HF}(u)(y,\underbrace{G_{LF}(u)(y)}_{\text{Augmented Input}})}_{\text{Learnable Residual}}
    \label{equation_MF}
\end{equation}
In addition, two other naive MF frameworks can also be used. One relies exclusively on residual learning, while the other solely augments the trunk input, as illustrated below:
\begin{equation}
\begin{aligned}
    G_{MF} &= \underbrace{G_{LF}(u)(y)}_{\text{Frozen Prediction}} + \underbrace{\delta_{HF}(u)(y)}_{\text{Learnable Residual}} \\
    \\
    G_{MF} &= \underbrace{G_{HF}(u)(y,G_{LF}(u)(y))}_{\text{Direct Learning}}
\end{aligned}
\end{equation}

In our present framework, we adopt Equation \ref{equation_MF}. The LF model provides an initial rough prediction and subsequently enters into the HF model trunk. The LF model can provide extensive feature learning, and then input it to HF model. The HF model can further perform residual learning to improve accuracy. Finally, the MF model we obtain is the addition of residual learned by the HF model and the prediction from the LF model.

\subsection{Bayesian Operator Learning}
\label{sec:BNN}
The stochastic nature of spinodal and the uncertainty in real world need to be represented. One way to generate uncertainty is to use Bayesian Neural Network (BNN) \cite{blundell2015weight}. Rather than finding best estimation for the network's weights via Maximum Likelihood Estimation, BNN seeks a posterior distribution over the weights \cite{blundell2015weight}. This is achieved through Variational Inference (VI) \cite{blundell2015weight}, which approximates the intractable true posterior with a simpler parameterized distribution (e.g., a Gaussian). The training objective thus has a new term, which is the Kullback-Leibler (KL) divergence between the approximate and true posterior. This training need to balance prediction accuracy (minimizing the error loss) with a regularization term that quantifies model uncertainty. The BNN allows the output to produce a probabilistic prediction. Therefore, BNN was widely used in uncertainty quantifying \cite{wilson2017reparameterization, wilson2018maximizing, linka2025discovering}.

Here, we introduce the BNN for our MF operator training model, as shown in Figure~\ref{Result_LF}. Leveraging its ability to quantify uncertainty, we aim to efficiently represent the stochastic nature of spinodal architectures. This model is built upon BNN-DeepONet, thus it is capable to provide both prediction mean and epistemic uncertainty estimates of stress–strain response. In addition, the training process of LF model \(G_{LF}\) is governed by a hybrid active learning strategy (discussed in details in next section) that iteratively selects the most informative design parameters from a large unlabeled pool.

Consider an operator $G$,  for a given input $u \in \mathcal{U}$, it will lead to an output $G(u)$. For the given coordinate $y \in \mathcal{Y}$, the DeepONet approximates the operator values at query location $G(u)(y)$ via the inner product of two neural network outputs:
\begin{equation}
    G(u)(y) \approx \sum_{k=1}^{p} \underbrace{b_k(u)}_{\text{Branch}} \cdot \underbrace{t_k(y)}_{\text{Trunk}} + b_0
\end{equation}
where $\{b_k\}_{k=1}^p$ and $\{t_k\}_{k=1}^p$ are the basis functions learned by the Branch and Trunk neural networks. They are p-dimensional latent space, respectively. In addition, $b_0$ is a bias term, which is optional but could reduce generalization error in some scenarios \cite{lu2021learning}. 

Given the training data $\mathcal{D}={\{u,X(y)\}}$, where $u$ is the input and $X(y)$ is the output at position $y$. We learn the operator mapping $G$ that give us $X^* = G(u^*)(y)$ by doing regression task. In standard neural networks, deterministic weights $\mathbf{w}(\mathcal{D})$ are optimized by minimizing loss function via backpropagation and gradient descent:
\begin{equation}
    \mathbf{w}=\underset{\mathbf{w}}{\operatorname{argmin}} \sum_*\left\|X^*-G^\mathbf{w}(u^*)(y^*)\right\|
\end{equation}
where $G^\mathbf{w}$ is the trained operator with deterministic weights $\mathbf{w}$.  Compared with optimizing deterministic weights $\mathbf{w}(\mathcal{D})$, BNN learns the posterior distribution of the weights $P(\mathbf{w}|\mathcal{D})$:
\begin{equation}
\begin{aligned}
    \mathbf{w} & =\arg \max _{\mathbf{w}} \log P(\mathcal{D} \mid \mathbf{w}) \\
    & =\arg \max _{\mathbf{w}} \sum_* \log P\left(X^*(y) \mid u^*, \mathbf{w}\right) 
\end{aligned}
\end{equation}
This indicates that each weight has a distribution. A simple representation is the Gaussian distribution, so that each weight $w_i$ has a mean $\mu_i$ and a variance $Var_i$:
\begin{equation}
    w_i \sim \mathcal{N}\left(\mu_i, Var_i^2\right)
\end{equation}
Thus, BNN can apply Variational Inference techniques to approximate the intractable true posterior with a parameterized variational distribution $q_\phi(\mathbf{w})$, typically a multivariate Gaussian. The training objective is to minimize the Evidence Lower Bound, which is equivalent to minimizing the combination of the negative log-likelihood and the KL divergence:
\begin{equation}
    \mathcal{L}(\phi) = \underbrace{\beta_{\text{KL}} \cdot \text{KL}[q_\phi(\mathbf{w}) || P(\mathbf{w})]}_{\text{Prior Regularization}} - \underbrace{\mathbb{E}_{q_\phi(\mathbf{w})}[\log P(\mathcal{D}|\mathbf{w})]}_{\text{Data Likelihood}}
    \label{loss_function}
\end{equation}
where $P(\mathbf{w})$ is the prior distribution (standard normal) and $\beta_{\text{KL}}$ is a weighting factor for balancing KL term (distribution difference) and Likelihood term (or MSE term). In this work, we utilize the "Flipout" estimator \cite{wen2018flipout} from the \texttt{bayesian-torch} library \cite{krishnan2022bayesian}, which decorrelates gradients within mini-batches to reduce variance and accelerate convergence. During inference, we can perform $T$ times Monte Carlo forward passes to approximate the predictive posterior
\begin{equation}
\begin{aligned}
    {\text{Mean}[G(u)(y)]} & \colonapprox \frac{1}{T}\sum_{t=1}^T G^\mathbf{{w}_t}(u)(y), \\
    {\text{Var}[G(u)(y)]} & \colonapprox \frac{1}{T}\sum_{t=1}^T (G^\mathbf{{w}_t}(u)(y) - \text{Mean}[G(u)(y)])^2
\end{aligned}
\end{equation}
where $\mathbf{w}_t \sim q_\phi(\mathbf{w})$ is randomly generated weights based on the trained distribution. By sampling $\mathbf{w}_t$ for a sufficient number of times, we can obtain the uncertainty quantification result.

\subsection{Hybrid Active Operator Learning for Low-Fidelity Model}
\label{sec:LF model}
In this work, we adopt a hybrid AOL strategies when training the LF model, which allows the strategical selection of the most informative unlabeled input function for labeling (to obtain the corresponding output), and then retrain the model until the operator learning reaches the halting criterion. This effective sampling can further reduce the sample size for training the surrogate model. This is especially crucial in domains where labeling is expensive or time-consuming, such as our spinodal microstructure and in-situ nanomechanical testing. 

\begin{figure*}[!htbp]
    \centering
    
    \includegraphics[width=1\linewidth]{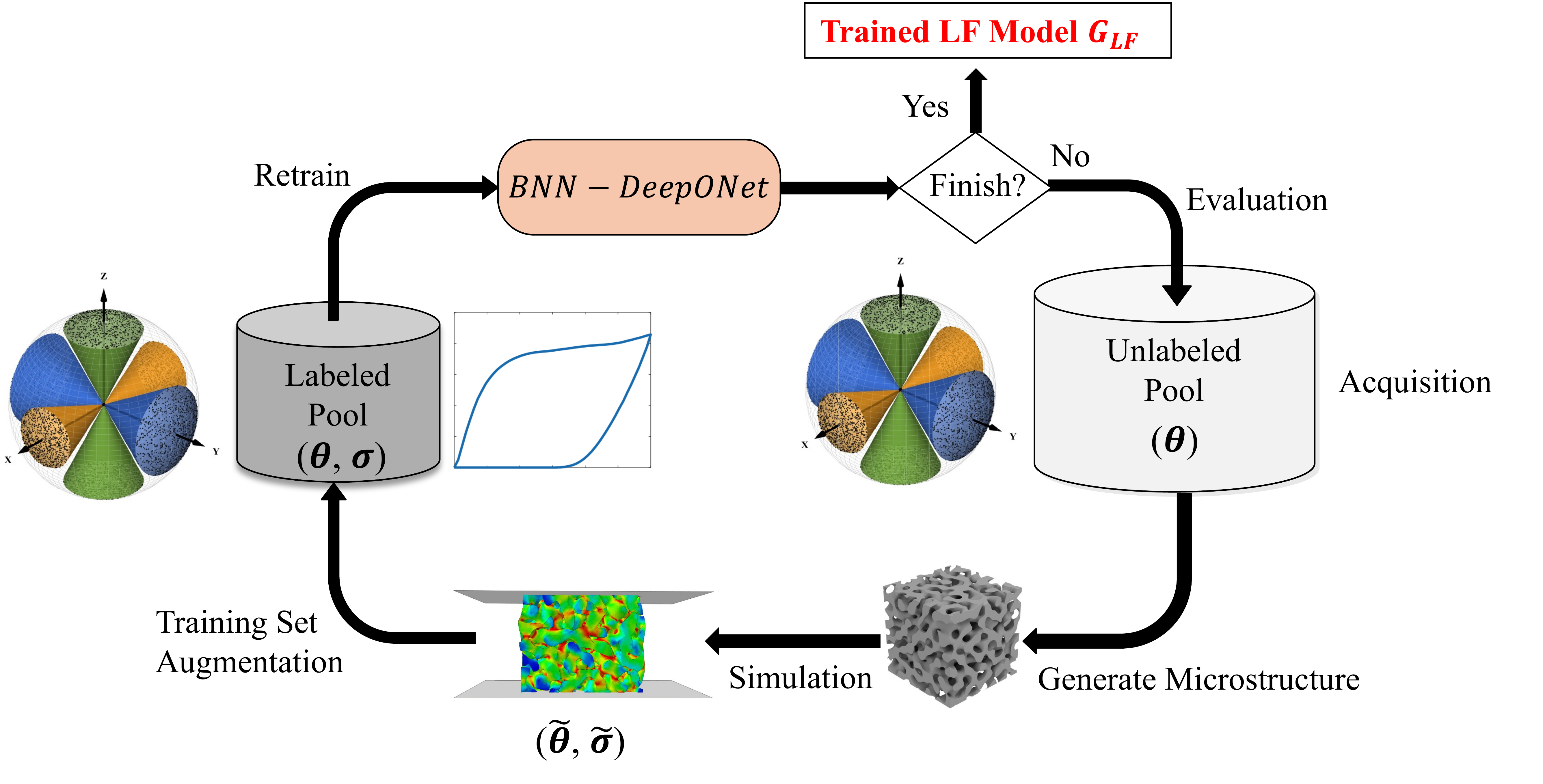}
    \caption{Workflow of the Hybrid Active Operator Learning for spinodal microstructure. Samples are selected from the pool based on a combined score of model uncertainty and geometric distance to the existing training set.}
    \label{AL_scheme}
\end{figure*}

Figure~\ref{AL_scheme} shows the workflow of AOL incorporated with LF model training. The key is to use an appropriate strategy to do acquisition based on the current model, then augment the training dataset, and finally retrain the model. Here, we adopt two metrics for AOL: uncertainty and diversity \cite{ren2021survey}. We employ this pool-based Hybrid AOL strategy for unlabeled sample acquisition which balances \textit{exploitation} (reducing model uncertainty) and \textit{exploration} (covering the whole design space). Specifically, for every candidate sample $x$ in the unlabeled pool $\mathcal{P}$, we compute a hybrid acquisition score $\alpha(x)$, which consists of two terms:

\textbf{1. Uncertainty Score ($\mathcal{S}_{var}$):} First we target samples where the model exhibits high epistemic uncertainty relative to the magnitude of the model prediction. This relative definition prevents the model from focusing on regions with naturally high stress simply because the absolute variance is higher, which could reduce the generalization ability. For a candidate input $\mathbf{u}$, the score is computed via $T=50$ Monte Carlo forward passes to get mean and variance, and their corresponded score:
\begin{equation}
    \mathcal{S}_{var}(\mathbf{u}) = \max_{\mathbf{y}} \left( \frac{\text{Var}[G(\mathbf{u})(\mathbf{y})]}{\max_{\mathbf{y}}|\mathbb{E}[G(\mathbf{u})(\mathbf{y})]| + \epsilon} \right)
\end{equation}
where $\epsilon=10^{-8}$ ensures numerical stability, ${\max_{\mathbf{y}}|\mathbb{E}[G(\mathbf{u})(\mathbf{y})]|}$ represents the magnitude of output function along all coordinate, and $\epsilon$ is a small value for robustness.

\textbf{2. Diversity Score ($\mathcal{S}_{dist}$):} To ensure the coverage of the whole topological design space, we compute the minimum-weighted Euclidean distance from a candidate $\mathbf{u}$ to the nearest neighbor in the existing training set $\mathcal{D}_{train}$:
\begin{equation}
    \mathcal{S}_{dist}(\mathbf{u}) = \lambda(\mathbf{u}) \cdot \min_{\mathbf{u}_j \in \mathcal{D}_{train}} \| \mathbf{u} - \mathbf{u}_j \|_2
\end{equation}
with
\begin{equation}
    \lambda(\mathbf{u})= \begin{cases}1 & \text { if } \mathbf{u} \in\left(0,0, \theta_1\right) \\ \frac{1}{\sqrt{2}} & \text { if } \mathbf{u} \in\left(0, \theta_2, \theta_1\right) \\ \frac{1}{\sqrt{3}} & \text { if } \mathbf{u} \in\left(\theta_3, \theta_2, \theta_1\right)\end{cases}
\end{equation}
where $\lambda(\mathbf{u})$ is a weighting term accounting for the invariant distance metric for diverse spinodal design angles which ensures the consistency of the distance metric across the three groups of samples, thus enabling more uniform acquisition.

Therefore, the final score is a sum of these metrics, normalized over the current labeled pool to range $[0, 1]$:
\begin{equation}
    \alpha(u) = \frac{\mathcal{S}_{var}(u) - \min(\mathcal{S}_{var})}{\max(\mathcal{S}_{var}) - \min(\mathcal{S}_{var})} + \frac{\mathcal{S}_{dist}(u) - \min(\mathcal{S}_{dist})}{\max(\mathcal{S}_{dist}) - \min(\mathcal{S}_{dist})}
\end{equation}

Since the model needs to be retrained after each time new samples were added, batch-sample active learning is used to reduce training times. At each step, we select a certain number of samples with a fixed batch size to add to the training set. It is makes sense to rank all sample for pool from high score to low score. However, computing the minimum distance of selected sample with training set (i.e., the second score) requires considering the possibility of the newly added samples on the distance calculation. Therefore, using this query strategy to find batch samples is an NP-hard problem \cite{ren2021survey}. In this consideration, we use a greedy strategy for solution. The acquisition process is detailed in Algorithm \ref{alg:hybrid_al}.

\begin{algorithm}
\caption{Hybrid Active Operator Learning Strategy}
\label{alg:hybrid_al}
\begin{algorithmic}[1]
\Require Unlabeled Pool $\mathcal{P}$, Initial Training Set $\mathcal{D}_{train}$, Max Steps $N$, Batch Size $B$
\Ensure Optimized Bayesian-DeepONet Model ${G}_{LF}$
\State Initialize ${G}_{LF}$ with random weights
\State Train ${G}_{LF}$ on $\mathcal{D}_{train}$
\For{step $n = 1$ to $N$}
    \State $\mathcal{S}_{var} \gets [], \mathcal{S}_{dist} \gets []$
    \For{each candidate $u_i$ in $\mathcal{P}$}
        \State \textbf{Predict} mean $\mu(u_i)$ and variance $Var^2(u_i)$ using ${G}_{LF}$ (Monte Carlo)
        \State \textbf{Compute} Uncertainty: $s_{v} \gets \max(Var^2 / (|\mu| + \epsilon))$
        \State \textbf{Compute} Diversity: $s_{d} \gets \min_{u_j \in \mathcal{D}_{train}} \| u_i - u_j \|$
        \State Append $s_{v}$ to $\mathcal{S}_{var}$, $s_{d}$ to $\mathcal{S}_{dist}$
    \EndFor
    \State \textbf{Normalize} $\mathcal{S}_{var}$ and $\mathcal{S}_{dist}$ to $[0, 1]$
    \State \textbf{Compute} Total Score: $\mathbf{\alpha} \gets \mathcal{S}_{var}^{norm} + \mathcal{S}_{dist}^{norm}$
    \State \textbf{Select} batch $\mathcal{B} \subset \mathcal{P}$ of size $B$ with highest scores $\mathbf{\alpha}$
    \State \textbf{Label} $\mathcal{B}$ via FEM Simulation (Oracle)
    \State $\mathcal{D}_{train} \gets \mathcal{D}_{train} \cup \mathcal{B}$
    \State $\mathcal{P} \gets \mathcal{P} \setminus \mathcal{B}$
    \State \textbf{Retrain} ${G}_{LF}$ on updated $\mathcal{D}_{train}$
\EndFor
\State \Return ${G}_{LF}$
\end{algorithmic}
\end{algorithm}

In this training step, the LF model \(G_{LF}\) approximates the operator that mapping the 3 design angles to the stress–strain response. The architecture details of two subnetworks are:
\begin{enumerate}
    \item \textbf{Branch Network:} A Multilayer Perceptron (MLP) encoding the input function space parameters $\mathbf{u} = [\theta_1, \theta_2, \theta_3]$. Based on our implementation, the branch net comprises an input layer, two hidden layers with 32 neurons each, and an output projection to a 128-dimensional latent space ($p=128$). Rectified Linear Unit (ReLU) activation functions and a dropout rate of 0.1 are applied after each hidden layer.
    \item \textbf{Trunk Network:} A MLP encoding the coordinate space $\mathbf{y} = [\varepsilon, \phi, d_x, d_y, d_z]$, where $\varepsilon$ is strain (range from 0 to 0.3), $\phi$ is loading phase (1 and 0 represent loading and unloading), and $\mathbf{d}$ is the loading direction vector ([1,0,0], [0,1,0] , [0,0,1] represent compression at X, Y, Z direction respectively). Similar to the branch net, the trunk net utilizes two hidden layers of 32 neurons and outputs a 128-dimensional latent vector.
\end{enumerate}

Besides, we employ VI by converting the standard deterministic layers into Bayesian layers using the ``Flipout'' estimator \cite{wen2018flipout}. The network weights $\mathbf{w}$ are treated as random variables parameterized by a posterior distribution $q_\phi(\mathbf{w})$. Based on this, the loss function is defined as:
\begin{equation}
    \mathcal{L}(\mathbf{w}) = \text{MSE}(\hat{\sigma}, \sigma_{L}) + \lambda_{KL} \cdot \text{KL}[q_\phi(\mathbf{w}) || P(\mathbf{w})]
\end{equation}
where $\hat{\sigma}$ is the ground truth stress of simulation while $\sigma_{L}$ is model's output stress, $P(\mathbf{w})$ is the standard normal prior, $\lambda_{KL}=10^{-5}$ is the weighting factor, and the MSE also represents the negative log-likelihood term under a Gaussian assumption. We utilize the Adam optimizer with a learning rate of $10^{-3}$ and train for 1,000 epochs per active learning iteration.

\subsection{SDF-based Residual Learning for High-Fidelity Model}
\label{sec:HF model}
While the LF model efficiently covers the design parameter space, it relies on idealized geometry and simplified material laws. HF experimental data captures the true physics, including manufacturing defects, surface roughness, and complex contact mechanics, but is extremely sparse. To bridge this reality gap, we implement the MF Residual Learning with augmented trunk framework.

The HF experimental response $\sigma$ was expressed as the superposition of the LF prediction $\sigma_{L}$ and a learnable residual term $\sigma_{\delta}$:
\begin{equation}
\begin{split}
    \sigma(\mathbf{u}, \mathbf{y}) & = \sigma_{L} + \sigma_{\delta} \\ 
    &  \approx \underbrace{G_{LF}(\mathbf{u}_{angles})(\mathbf{y})}_{\text{LF Prediction } \sigma_{L}} + \underbrace{G_{HF}(\mathbf{u}_{SDF})({\sigma_{L},\mathbf y})}_{\text{HF Learned Residual } \delta_{residual}}
\end{split}
\end{equation}
where $\mathbf{u}={\mathbf{u}_{angles} \oplus \mathbf{u}_{SDF}}$ is the geometric input (i.e. design angles and SDF of microstructure) with $\mathbf{y}$ is query coordinates, $G_{LF}$ is the stage-1 trained LF operator model driven by the design angles $\mathbf{u}_{angles}$, and $G_{HF}$ is the residual operator learning using HF data. Crucially, the residual is often driven by geometric deviations not captured by the three design angles (e.g., local strut variations or printing artifacts). Therefore, for the input function for the residual model we adopt the full 3D geometry of the printing model of samples, represented as a SDF, $\mathbf{u}_{SDF}$. Here the SDF is the geometrical representation by storing the shortest distance from uniform grid point in the cube to that microstructure's surface, which can be computed using many open-source libraries, such as \cite{park2019deepsdf}.

To process the high-dimensional volumetric data SDF, we replace the standard MLP branch network with a 3D CNN, as shown in Figure \ref{Whole_architecture}. The HF architecture comprises:
\begin{itemize}
    \item \textbf{SDF Branch Net:} The input is a grid saved SDF of size $41 \times 41 \times 41$. The network acts as a feature extractor using two blocks of 3D convolution and max-pooling:
    \begin{enumerate}
        \item Conv3D ($1 \to 4$ channels, $3\times3\times3$ kernel) $\to$ ReLU $\to$ MaxPool3D ($2\times$)
        \item Conv3D ($4 \to 16$ channels, $3\times3\times3$ kernel) $\to$ ReLU $\to$ MaxPool3D ($2\times$)
    \end{enumerate}
    After these two layers, the flatten layer with two fully connected layers was used to map the resulting feature into the 128-dimensional latent space.
    \item \textbf{Trunk Net:} The 3 inputs are identical to the LF model, while the newly augmented one is LF model predicted stress. It accepts the coordinate vector $\mathbf{y} = [\sigma_{L},\varepsilon, \phi, \mathbf{d}]$ and passes two hidden layers with 32 neurons  to output a 128-dimensional latent vector.
\end{itemize}

The residual model is trained on the discrepancy between the experiment and the LF mean prediction: $\delta_{residual} = \sigma_{exp} - \sigma_{L}$. We employ the same Bayesian formulation as the LF model, utilizing Flipout layers for uncertainty quantification. The loss function includes a KL divergence term (weight $\lambda_{KL}=10^{-5}$) and an MSE term on the residual. The learning rate is set to $10^{-3}$ with a scheduler that reduces the rate by a factor of 0.3 if loss does not reduce in the previous 50 epochs.

During inference, the final probabilistic prediction combines the deterministic mean of the LF baseline with the posterior predictive distribution of the residual:
\begin{equation}
    P(\sigma_{MF}) \sim \mathcal{N}(\mu({\sigma_{L}}) + \mu({\delta_{residual}}), Var({\delta_{residual}})^2)
\end{equation}
In short, the role of LF model \(G_{LF}\) is to explore diverse stress behaviors from FEM simulation, and thus it can provide a rough estimation. LF model \(G_{LF}\) do not need to achieve perfect accuracy, since the role of HF model \(G_{HF}\) is to learn the gaps $\delta_{residual}$ between simulation results and experimental results. This architecture ensures that the model retains the global trends learned from the massive simulation dataset while locally correcting for the reality gap using the sparse, geometry-aware experimental data.

\subsection{Loss Function Definition}
\label{sec:Loss functions}
The loss function terms used in the training process, as shown in Equation \ref{loss_function}, consists of two terms for accounting output accuracy and weight's  probability distribution: mean squared error (MSE) and Kullback–Leibler (KL) divergence. The MSE is defined as
\begin{equation}
    MSE = \frac{1}{303n}
    \sum_{k=1}^{n}
    \sum_{j=1}^{3}
    \sum_{i=1}^{101}
    \left( \hat{\sigma}_{ij}^{k} - {\sigma}_{ij}^{k} \right)^{2}
\end{equation}
where $\hat{\sigma}_{ij}^{k}$ and ${\sigma}_{ij}^{k}$ represent the ground truth (FEM data in LF model or experiment data in HF model) and model's prediction on $k$-th sample on the stress–strain curve at the $j$-th direction and the $i$-th loading point. $n$ is the total number of samples in training.

For the BNN where the weights are modeled as independent Gaussian distributions, the total KL divergence between the variational posterior $q(\mathbf{w})$ and the prior $p(\mathbf{w})$ is the sum of the KL divergences of individual weights. Considering the prior for each weight is $p(\mathbf{w}) = \mathcal{N}(\mu_p, Var_p^2)$ and the variational posterior be $q(\mathbf{w}) = \mathcal{N}(\mu_q, Var_q^2)$. The analytical form is:

\begin{equation}
    {KL}(q(\mathbf{w}) \parallel p(\mathbf{w})) = \sum_{i=1}^{N} \left[ \log \left( \frac{Var_p}{Var_{q,i}} \right) + \frac{Var_{q,i}^2 + (\mu_{q,i} - \mu_p)^2}{2 Var_p^2} - \frac{1}{2} \right]
\end{equation}
where N is the total number of weights in networks. In the implementation, the prior is set as a standard normal distribution where $\mu_p = 0$ and $Var_p = 1$. The equation simplifies to:
\begin{equation}
    {KL}(q(\mathbf{w}) \parallel p(\mathbf{w})) = \sum_{i=1}^{N} \frac{1}{2} \left( -\log(Var_{q,i}^2) + Var_{q,i}^2 + \mu_{q,i}^2 - 1 \right)
\end{equation}

\subsection{Metrics for Result Comparison}
To compare the training performance, other than loss functions, we introduce two more metrics. To assess the accuracy of the surrogate model in predicting unseen data, R-squared (${R}^2$) is also used. It indicates the model's goodness-of-fit, defined as:
\begin{equation}
    {R}^2 = 1- \frac{\sum_{k=1}^{n}
    \sum_{j=1}^{3}
    \sum_{i=1}^{101}
    \left( \hat{\sigma}_{ij}^{k} - {\sigma}_{ij}^{k} \right)^{2}}
    {\sum_{k=1}^{n}
    \sum_{j=1}^{3}
    \sum_{i=1}^{101}
    \left( \hat{\sigma}_{ij}^{k} - \bar{\sigma}_{ij}^{k} \right)^{2}}
\end{equation}
where $\hat{\sigma}_{ij}^{k}$ is the ground truth, ${\sigma}_{ij}^{k}$ is the model's prediction, and $\bar{\sigma}_{ij}^{k}$ is the mean of ground truth.

To assess uncertainty, the metric we used is the Prediction Interval Coverage Probability (PICP). Its formula is
\begin{equation}
\label{PICP}
    PICP = \frac{1}{303n}
    \sum_{k=1}^{n}
    \sum_{j=1}^{3}
    \sum_{i=1}^{101}
    \omega^{k}_{ij},
    \qquad
    \text{with } 
    \omega^{k}_{ij} =
    \begin{cases}
        1, & \text{if } \sigma^{k}_{ij} \in \{\hat{\sigma}^{k}_{ij} \pm 1.96\, \mathrm{Var}(\hat{\sigma}^{k}_{ij})\}, \\
        0, & \text{otherwise}.
    \end{cases}
\end{equation}
PICP quantifies the confidence score of BNN uncertainty prediction by counting the ratio of true values that fall within the 95\%-confidence interval of BNN prediction.

\section*{Contributions}
H.D.E. and B.B. conceived the research and managed the project. P.Y. performed the FE simulations, and H.C. organized the in-situ experiment data. P.Y. developed the neural network model. P.Y. wrote the initial manuscript. All authors contributed to the review and editing of the final manuscript.

\section*{Acknowledgments}
H.D.E. acknowledges financial support from the Office of Naval Research (grant N000142212133) and the Air Force Office of Scientific Research (Grant No. FA9550-23-1-0284). This research was supported in part through the computational resources and staff contributions provided for the Quest high-performance computing facility at Northwestern University, which is jointly supported by the Office of the Provost, the Office for Research, and Northwestern University Information Technology. The authors also acknowledge Hanxun Jin and Boyu Zhang, members of the Espinosa group at Northwestern University, for fabricating the samples and performing the in-situ SEM experiments.

\section*{Competing interests}
The authors declare no competing interests.

\section*{Data Availability}
Raw data were generated in our laboratory, and the methods are presented within this article and supplementary information. Derived data and additional information are available from the corresponding authors upon request.

\section*{Declaration on the Use of Generative AI}
The authors used ChatGPT, a large language model developed by OpenAI, to assist with English language editing, grammar refinement, and to provide guidance on LaTeX scripts. The scientific content, technical interpretations, and conclusions are solely the responsibility of the authors.

\renewcommand{\thefigure}{A\arabic{figure}}
\setcounter{figure}{0}

\bibliographystyle{unsrtnat}
\bibliography{bibliography}

\end{document}